\documentclass[twocolumn,secnumarabic,amssymb, nobibnotes, aps, prd]{revtex4}
\usepackage{graphicx}

\begin{document}
\title{Multiple superconducting ring ratchets for ultrasensitive detection of non-equilibrium  noise.}
\author{V.L. Gurtovoi$^{1,2}$, M. Exarchos$^{3}$, V.N. Antonov$^{3}$, A.V. Nikulov$^{1}$, and V.A. Tulin$^{1}$}
%\email[]{nikulov@ipmt-hpm.ac.ru}
\affiliation{$^{1}$Institute of Microelectronics Technology and High Purity Materials, Russian Academy of Sciences, 142432 Chernogolovka, Moscow District, Russia. \\ $^{2}$Moscow Institute of Physics and Technology, 29 Institutskiy per., 141700 Dolgoprudny, Moscow Region, Russia. \\ $^{3}$Physics Department, Royal Holloway University of London, Egham, Surrey
TW20 0EX, UK
} %nikulov@ipmt-hpm.ac.ru
%\date{}
\begin{abstract} Magnetic quantum periodicity in the dc voltage is observed when asymmetric rings are switched between superconducting and normal state by a noise or ac current. This quantum effect is used for detection of a non-equilibrium noise with the help of a system of 667 asymmetric aluminum rings of $1 \ \mu m$ in diameter connected in series. Any noise down to the equilibrium one can be detected with the help of such system with enough great number of asymmetric rings.
\end{abstract}

\maketitle 

\narrowtext

\section{Introduction}
Some authors \cite{ArtAt2010,ArtAt2011,ArtAt2012} consider superconducting loop as an artificial atom because its spectrum of permitted states is discrete due to the quantization postulated by Bohr's as far back as 1913 in order to describe stable electron orbits of atom.  These artificial atoms provide additional experimental opportunities for studies of quantization phenomena. The spectrum of electrons in atom is always discrete whereas the spectrum of the mobile charge carriers is discrete only in superconducting state. Therefore the transition from normal to superconducting state of a mesoscopic or macroscopic loop is the transition from continuous to discrete spectrum of permitted states of the mobile charge carriers. This transition is observed when whole ring \cite{PRB2014C} or its segment \cite{JLTP1998,PLA2012QF} is switched in superconducting state. 

According to the orthodox definition of the operator of canonical momentum $\hat{P} = -i\hbar \nabla $ \cite{LandauL} the total angular momentum of superconducting condensate homogeneous in a ring with a radius $r$ and a section $s$ should be equal 
$$M _{p} =sr|\Psi |^{2}\hbar \oint _{l} dl \nabla \varphi =  N _{s}\hbar n \eqno{(1)}$$
to product of the number of superconducting pairs in the ring $ N _{s} = |\Psi |^{2}s2\pi r$ and the angular momentum of each pair $m _{p} =  \hbar n $. The quantization takes place when all ring segments are in superconducting state because it is a consequence of the requirement $\oint _{l} dl \nabla \varphi = n2\pi $ that the complex wave function must be single-valued at any point of the ring $\Psi = |\Psi |e^{i\varphi } =  |\Psi |e^{i(\varphi + n2\pi )}$. The operator of canonical momentum is the same $\hat{P} = -i\hbar \nabla $ with and without magnetic field \cite{LandauL} whereas the operator of the velocity of a particle with a charge $q$ is $\hat{v} = (\hat{P} - qA)/m$ \cite{FeynmanL} changes in the presence of a magnetic vector potential $A$. Therefore the velocity $v$ and the superconducting current $I_{p} = sq|\Psi |^{2}v$ 
$$I_{p} = \frac{sq}{m2\pi r}\oint_{l}dl \Psi ^{*}(-i\hbar \nabla  - qA)\Psi  = \frac{n\Phi_{0} - \Phi}{L_{k}}  \eqno{(2)}$$
can not be equal zero when the magnetic flux inside the ring $\Phi = \oint_{l}dl A$ is not divisible $\Phi \neq n\Phi _{0}$ by the flux quantum $\Phi _{0} = 2\pi \hbar /q$. Here $L_{k} = ml/sq^{2}|\Psi |^{2}$ is the kinetic inductance of the ring with the length $l = 2\pi r$, the section $s$, the density of superconducting pairs $|\Psi |^{2}$ and the pair charge $q = 2e$. 

\begin{figure}[b]
\includegraphics{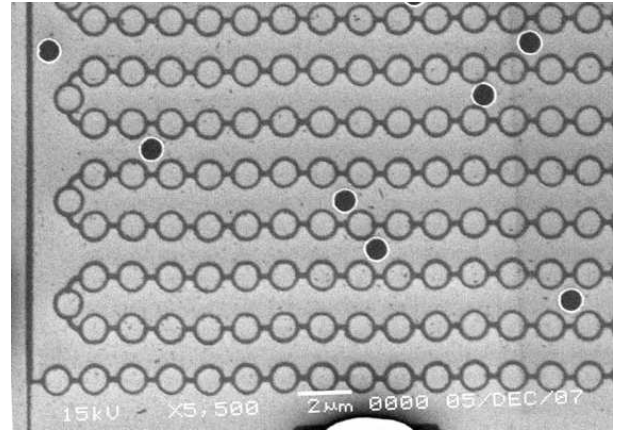}
\caption{\label{fig:epsart} A fragment of structure consisting of 667 asymmetric rings with diameter of $1 \ \mu m$.}
\end{figure} 

 The current (2) should appear when all ring segment are switched in superconducting state and should decay because of dissipation when at least one segment is switched in the normal state with a resistance $R$ \cite{JLTP1998,PLA2012QF}. Numerous measurements \cite{PCScien07,PCJETP07,JETP07J} testify to the predominant probability $P _{n} \propto \exp{- E_{n}/ k _{B}T} $ of the permitted state $n$ (1) with the minimal value of the kinetic energy 
$$E _{k} = \frac{1}{2m}\int _{V}dV\Psi ^{*} (-i\hbar \nabla  - qA)^{2}\Psi = \frac{(n\Phi_{0} - \Phi )^{2}}{2L_{k}}  \eqno{(3)}$$
The measurements of the critical current \cite{PCJETP07,JETP07J} testify that two values of  the persistent current (2) corresponding to $n$ and $n+1$ are observed only in a special case \cite{2state2005LT} because of the strong discreteness of spectrum (3): $(n\Phi_{0} - \Phi )^{2}/2L_{k} = (I _{p}\Phi_{0}/2)(n -  \Phi/\Phi_{0})^{2} \approx k _{B}80 \ K(n -  \Phi/\Phi_{0})^{2}$ at a typical value $ I _{p} \approx  1 \ \mu A$. The persistent current (2) should have the same direction with the predominant probability $P _{n}(\Phi)/ P _{n+1}(\Phi)  \approx  \exp {40} \approx  10^{17}$ at a given value $\Phi \approx  (n+0.25) \Phi_{0}$ and the temperature $T \approx  1 \ K$. The potential voltage with a direct component $V _{dc}$ may be expected to observe on the segment when it is switched between superconducting and normal states \cite{JLTP1998,PLA2012QF} due to this ratchet effect. The dc voltage should oscillate with magnetic field likewise the average value of the persistent current $ \overline{I_{p}} = (\overline{n}\Phi_{0} - \Phi )/L _{k}$, where $\overline{n} = \sum _{n}nP _{n}(\Phi)$ and $P _{n}(\Phi)$ is the probability of the switching in superconducting state with the quantum number $n$ at magnetic flux inside the ring $\Phi $. Similar oscillations were observed at the switching of asymmetric ring between superconducting and normal states induced by the ac current \cite{PCJETP07,Letter2003} or a noise \cite{PerMob2001,Letter07,toKulik2010,PLA2012}. These experimental evidence of the ratchet effect allows to use a system of asymmetric superconducting ring for ultrasensitive detection of non-equilibrium  noise. We demonstrate this opportunity on the example of a system of aluminum rings connected in series.

\begin{figure}[b]
\includegraphics{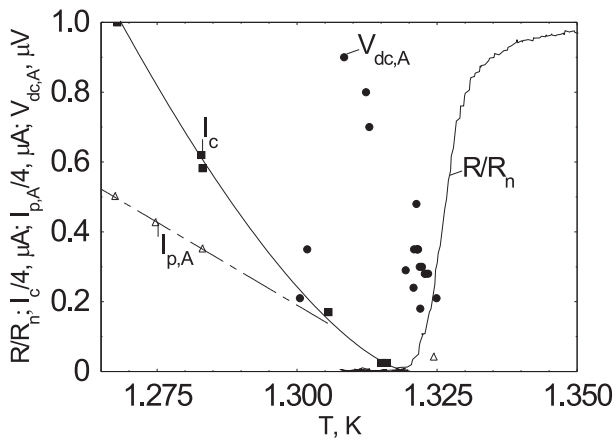}
\caption{\label{fig:epsart}  The superconducting resistive transition $R/R _{n}$, the temperature dependence of the critical current $I _{c}$ (squares, the line is the theoretical dependence), of the amplitude of the persistent current $I _{p,A}$ (triangles, the straight line is the theoretical dependence) and of the amplitude $V _{dc,A}$ (circles) of the dc voltage induced by an undesirable noise measured on the 667 rings.}
\end{figure} 

\section{Experimental Details}
In this work a structure with 667 asymmetric $1 \ \mu m$ in diameter rings, Fig. 1, were fabricated by e-beam lithography and lift-off process of 30 nm thick aluminum film. Ring arm widths were 100 and 125 nm for narrow and wide parts, respectively. For this structure, the resistance in the normal state $R _{n} = 5400 \ \Omega$, the resistance ratio R(300 K)/R(4.2 K) = 2, superconducting transition temperature $T _{c}$ was 1.320 K and the width of resistive transition $\Delta T_ {c}$ was 0.009 K, Fig.2. The temperature dependence of the critical current measured without magnetic field is described by the theoretical relation $I _{c} = I _{c}(T=0)(1 - T/T _{c})^{3/2}$, where $I _{c}(T=0) \approx 520 \ \mu A$, Fig.2. The critical current density $j _{c}(T=0) \approx 10^{7} A/cm^{2}$ equals approximately the depairing current density \cite{PCJETP07}. Measurements were carried out by applying DC, sinusoidal or white noise (bandwidth from DC to 200 kHz) bias current to current leads from ultra low distortion generator with differential output (Stanford Research, Model DS360) whereas, voltage corresponding to rectified, Little-Parks, R(T) or IV signal was measured in a frequency band from 0 to 30 Hz by an instrumentation amplifier (followed by a low-noise preamplifier SR560) at potential leads. Noise level of the amplification system was 20 nVpp for $f _{b}=0$ to 1 Hz. It should be noted that rectification effects do not depend on frequency of the bias current at least up to 1 MHz. Magnetic field direction was perpendicular to the ring's plane. Magnetic field time scanning was slow enough (~ 0.001-0.1 Hz) so that upper frequency of signal spectrum, which resulted from magnetic field changes, did not exceed 30 Hz. All signals corresponding to rectified voltage, current, temperature and magnetic field were digitized by an 8-channel 16-bit analog-to-digital converter card. 

\begin{figure}[b]
\includegraphics{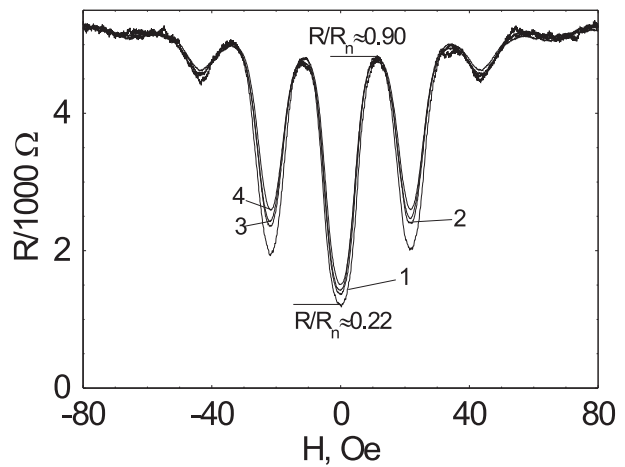}
\caption{\label{fig:epsart}  The resistive oscillations (the Little-Parks effect) of the 667 rings measured at different values of the measuring current and temperature corresponding to the resistive transition: 1) $I_{m} = 1 \ nA $, $T \approx  1.3244 \ K$; 2) $I_{m} = 2 \ A $, $T \approx  1.3253 \ K$; 3) $I_{m} = 3 \ nA $, $T \approx  1.3254 \ K$; 4) $I_{m} = 5 \ nA $, $T \approx  1.3257 \ K$. The period of oscillations $B _{0} = \Phi _{0}/S \approx  22 \ Oe$ corresponds to the ring's area $S = \pi r^{2} \approx  0.94 \ \mu m^{2}$ and $r \approx 0.55 \ \mu m$ .}
\end{figure} 

The magnetic field dependences of the critical currents $I_{c+}(B)$ and $I_{c-}(B)$ were determined by measuring periodically repeating current-voltage characteristics (a period of 10 Hz) in a slowly varying magnetic field $B_{sol}$ (a period of approximately 0.01 Hz) as follows. First, the condition that the structure was in the superconducting state was checked. Next, after the threshold voltage was exceeded (this voltage, set above induced voltages and noises of the measuring system, determined the minimum measurable critical current), magnetic field and critical current (with a delay of about $30 \ \mu s$) were switched on. This procedure allowed us to measure sequentially critical currents in the positive $I_{c+}$  and negative $I_{c-}$  directions with respect to the external measuring current $I_{ext}$. Measurements of one $I_{c+}(B)$ or $I_{c-}(B)$  dependence (1000 values) took about 100 s. Little-Parks oscillations $R(B) = V(B)/I_{ext}$ were recorded at a constant $I_{ext} = 0.1 \div 2.0 \ \mu A$ current. The field dependences of rectified voltage $V_{dc}(B)$ were measured using sinusoidal current $I_{ext}(t) = I_{0}\sin (2\pi ft)$ with the amplitude $I_{0}$ up to $50 \ \mu A$ and frequency $f = 0.5 \div 5 \ kHz$. Because of incomplete screening, the minima of the $R(B_{sol})$ solenoid field dependences of resistance and zero rectified voltage $V_{dc}(B_{sol})$ were shifted by $-B_{res} \approx  -0.15 \ G$. The $R(B_{sol} + B_{res})$ dependences had minima at $B_{sol} + B_{res} = n\Phi _{0}/S$ and maxima at $B_{sol} + B_{res} = (n+0.5)\Phi _{0}/S$ and the $V_{dc}(B_{sol} + B_{res})$ dependences intersected zero at these values of the total magnetic field $B_{sol} + B_{res}$. Because the simultaneous reversal of the total external field {\bf B} and measuring current $I_{ext}$ was equivalent to rotation through $180^{o}$, the equality $I_{c+}(B) = I_{c-}(-B)$ had to be satisfied. We had $I_{c+}(B_{sol} + B_{res}) = I_{c-}(-B_{sol} - B_{res})$ for all the dependences measured. This proves that $B_{sol} + B_{res}$ was the total external field also in critical current measurements.

\begin{figure}
\includegraphics{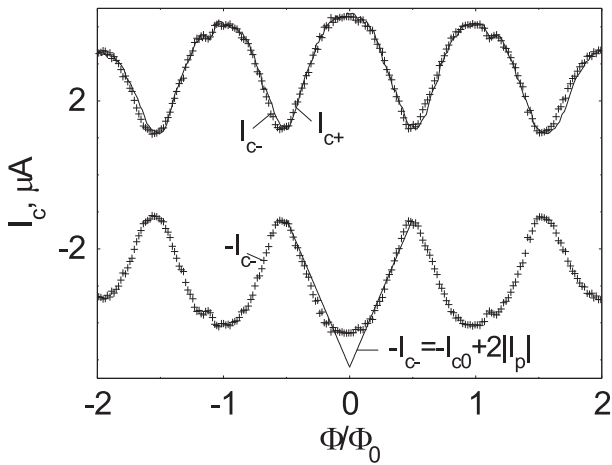}
\caption{\label{fig:epsart}  Magnetic dependence of the critical current of 667 rings measured at the temperature $T = 1.2675 \ K$ in the opposite directions $I _{c+}(\Phi /\Phi _{0})$ (line) and $I _{c-}(\Phi /\Phi _{0})$ (criss-crosses). The experimental dependence $-I _{c-}(\Phi /\Phi _{0})$ is compared with the theoretical one (4). The values $I _{c0} = 5.2 \ \mu A$ and $I _{p,A} = 2 \ \mu A$ were used for the theoretical dependence. }
\end{figure}

\section{Properties of the 667 rings system}  
The observation of the resistive oscillations (the Little-Parks effect \cite{LP1962}) in the temperature region $T > T _{c}$ testify to the existence of the persistent current in the fluctuation region, above superconducting state. The resistance increase from $R/R _{n} \approx  0.22$ at $B = 0$ to $R/R _{n} \approx 0.90$ at $B \approx \Phi _{0}/2S$, Fig.3, corresponds to the critical temperature decrease (the shift of the resistive transition) on $\Delta T _{c} \approx -0.01 \ K$. The results of our measurements of the resistive transition made at different value of the measuring current, from $I_{m} = 1 \ nA $ to $I_{m} = 500 \ nA $, allow to find that this shift corresponds to  $I_{m} \approx  300 \ nA $. Consequently the amplitude of the persistent current oscillations $I_{p,A} \approx  150 \ nA $ at $T \approx 1.325 \ K > T _{c} \approx 1.320 \ K$ exceeds the value of $I_{m} = 1 \div  5 \ nA $ used for the observation of the Little-Parks oscillations shown on Fig.3.

\begin{figure}[b]
\includegraphics{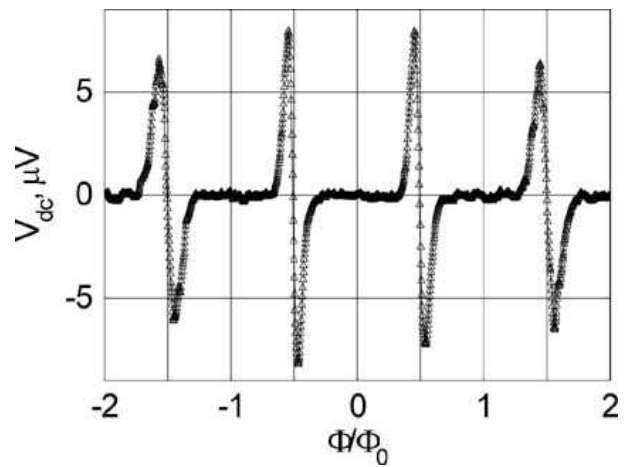}
\caption{\label{fig:epsart}  The quantum oscillations of the dc voltage induced by an electric noise with the amplitude 150 nA at the temperature $T = 1.307 \ K$. }
\end{figure}

The persistent current (2) results to the oscillations of the critical current measured in superconducting state at $T < T _{c}$, Fig.4. Magnetic dependence of the critical current of a symmetric ring may be described by the relation  
$$I _{c} = I _{c0}-2| I _{p}| = I _{c0}-2I _{p,A}2|n - \frac{\Phi }{\Phi _{0}}| \eqno{(4)}$$
See the deduction of the relation (4) in \cite{PCJETP07}. This relation describes enough well the experimental dependence of the critical current $I _{c+}$, $I _{c-}$ of 667 rings measured in opposite directions, Fig.4. The discrepancy between theoretical and experimental values near  $\Phi = 0$ may be explained the influence of the contacts between rings the width of which is smaller than the total width of two ring-halves, see Fig.1. The comparison of the theoretical and experimental dependence allows to find the amplitude $I _{p,A}(T)$  of the persistent current oscillation at different temperatures. The experimental temperature dependence may be described in the superconducting state by the theoretical relation  $I _{p,A} = I _{p,A}(T=0)(1 - T/T _{c})$, where $I _{p,A}(T=0) \approx  50 \ \mu A$, Fig.2. The amplitude should be equal zero at $T > T _{c} = 1.320 \ K$. But the Little-Parks oscillations, Fig.3, testify to a non-zero value of  $I _{p,A}$ at $T > T _{c}$, Fig.2, due to thermal fluctuations. 

\begin{figure}
\includegraphics{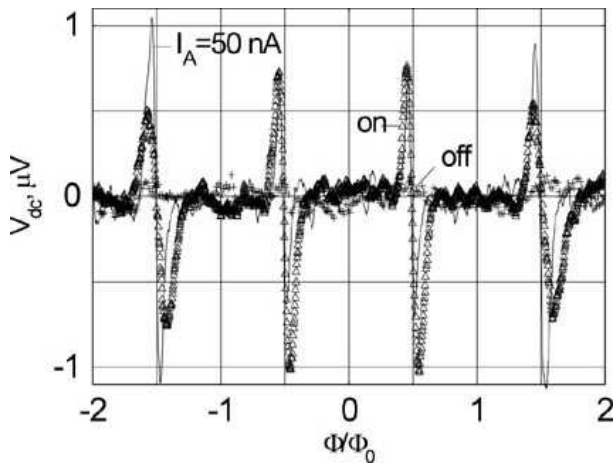}
\caption{\label{fig:epsart}  Magnetic dependence of the dc voltage measured on 667 rings at $T = 1.307 \ K$ when devices Nanovoltmeter Keithley 2182A and Current Sour Keithley 6221 were switched off (criss-crosses) and switched on (triangles) and when the dc voltage $V _{dc}(\Phi /\Phi _{0})$ (line) is induced by controllable noise with the amplitude $I _{A} = 50 \ nA$.}
\end{figure}

The critical current of symmetric rings measured in opposite directions equals each other $I _{c+} = I _{c-}$ and $I _{c+} - I _{c-} = 0$. The theory predicts that critical currents of asymmetric ring with different section of the ring-halves should be different $I _{c+} \neq I _{c-}$ and the anisotropy of the critical current $ I _{c,an} = I _{c+} - I _{c-}$ should be observed \cite{PCJETP07}. The anisotropy is indeed observed, see Fig.10, 16 in \cite{PCJETP07}. But its cause differs from the expected one, see Fig. 19 in \cite{PCJETP07}. The anisotropy $ I _{c,an}$ appears due to the shift of the dependence $I _{c+}(\Phi /\Phi _{0})$, $I _{c-}(\Phi /\Phi _{0})$ in the opposite direction \cite{JETP07J}. The shift arrives at a quarter of the flux quantum $ \Phi_{0}/4$ \cite{ PCJETP07,JETP07J}. The shift of the dependence $I _{c+}(\Phi /\Phi _{0})$, $I _{c-}(\Phi /\Phi _{0})$ of the critical current of the 667 rings with $r \approx  0.5 \ \mu m$ is hardly noticeable in comparison with the one of the single ring with $r \approx  2 \ \mu m$ and more strong asymmetry \cite{ PCJETP07,JETP07J}. The comparison of the magnetic dependence $I _{c+}(\Phi /\Phi _{0})$ and $I _{c-}(\Phi /\Phi _{0})$ demonstrates very slight anisotropy of the critical current, Fig.4. Nevertheless the rectification effect is observed also on the system of the 667 rings, Fig.5. The observation of the rectified voltage only in the regions of the $\Phi $ values near $\Phi = (n+0.5)\Phi _{0}$ corresponds to the observation of the critical current anisotropy $ I _{c,an}$ only in these regions, see Fig.4. 

\section{Detection of an undesirable noise created by a noisy equipment }  
A precise detection of any undesirable noise is needed for some purposes. Some precision devices, for example quantum bits, superconducting quantum interference device and others,  can not effectively work without precision control of external and internal electric noise. Results of measurements presented on Fig.6 demonstrate that a system of asymmetric superconducting rings connected in series can be effectively used for the detection of a undesirable noise. The amplitude of the dc voltage near $\Phi \approx (n+0.5)\Phi _{0}$ does not exceed $0.1 \ \mu V$ when devices Nanovoltmeter Keithley 2182A and Current Sour Keithley 6221, used for temperature measurement, were switched off and increases up to $1 \ \mu V$  when the devices are switched on, Fig.6. These observations give evidence that these devices creates an undesirable noise. The system of asymmetric superconducting rings may be calibrated with the help of a controllable noise, when the noisy equipment is switched off. The controllable noise with the amplitude $I _{A} \approx 50 \ nA$ induces approximately the same value of the dc voltage, as the noisy devices, Fig.6. We may conclude that the noisy devices induce a noise with the amplitude $I _{A} \approx 50 \ nA$ in the system of the 667 rings. 

The asymmetric superconducting ring is a ratchet because of the predominate probability of one of the directions of the persistent current. The dc voltage oscillations $V _{dc}(\Phi /\Phi _{0})$ are observed when the persistent current exists and when the rings are switched between superconducting and normal state. Their amplitude $V _{dc,A}$ decreases with the temperature increase at $T > 1.308 \ K$ because of the decrease of the amplitude of the persistent current $I _{p,A}$, Fig.2. The persistent current amplitude increases with the temperature decrease, Fig.2. But the critical current increases also. The non-equilibrium noise can not switch the rings in the normal state when the minimal value of the critical current (4) exceeds its amplitude $I _{A} \approx 50 \ nA$. Therefore the visible oscillations $V _{dc}(\Phi /\Phi _{0})$ with the amplitude $V _{dc,A} \geq  0.2 \ \mu V$ are observed only in a narrow temperature region, Fig.2. This region expands with the increase of the noise amplitude  $I _{A} $ and diminishes down to zero with the $I _{A} $ decrease. We can not detect the noise with the amplitude $I _{A} < 10 \ nA$ because the oscillations $V _{dc}(\Phi /\Phi _{0})$ with the amplitude $V _{dc,A} < 0.2 \ \mu V$ are not visible for us. In order to detect the smaller noise a system with larger number of asymmetric superconducting rings should be used. 

The dc voltage of a system of asymmetric superconducting rings connected in series increases in proportion to the number of rings \cite{Letter2003}, as well in series of any dc power sources. The amplitude $V _{dc,A}$ of the oscillations $V _{dc}(\Phi /\Phi _{0})$ induced by the same noise increases with the number $N$ of rings connected in series. Therefore the amplitude $I _{A} $ of the detectable noise decreases with the $N$ increase \cite{PLA2012}. The noise with the amplitude $ I _{A} > V _{A,vis}/R _{n} Eff _{Re}$ can be detected with the help of a ring system with the resistance in the normal state $R _{n} = NR _{n,1}$ and the maximum rectification efficiency  $Eff _{Re} = V _{A,max}/R _{n}I _{A}$ when the oscillations $V _{dc}(\Phi /\Phi _{0})$ with the amplitude $V _{A,vis}$ are visible. The rectification efficiency of asymmetric aluminium rings is abnormally high $Eff _{Re} \approx  0.33$ in the superconducting state at $T < 0.98T _{c}$ because of the hysteresis of the current - voltage characteristics in this temperature region \cite{PCJETP07}. A noise with the amplitude down to $ I _{A}  \approx 0.1 \ nA = 10^{-10} \ A$ could be detected with the help of the 667 ring system with $R_{n} \approx  5400 \ \Omega $ and this value of the rectification efficiency when $V _{A,vis} \approx  0.2 \ \mu V$. But the rectification efficiency decreases at $T \rightarrow T _{c}$ \cite{PCJETP07} and at $I _{A } \rightarrow 0$ \cite {PLA2012}. Therefore the noise with the amplitude $ I _{A}  > 10 \ nA$ could be detected with the help of our system of 667 rings. 

In order the rectification efficiency  $Eff _{Re} = V _{dc,A}/R _{n}I _{A} = \sum_{i=1}^{i=N}V _{dc,i}(B,T)/R _{n}I _{A}$ could have a maximum value the dc voltage $V _{i}(B,T)$ induced by the noise on each ring $i$ should have a maximum value at the same externally produced magnetic field $B$ and the temperature $T$. Therefore all rings should have the same area $S = \pi r^{2}$, because $\Phi = BS$, and the same critical temperature $T _{c}$, because the amplitude $V _{dc,A}$ has maximum at a $T/T _{c}$ value, Fig.2. It is not difficult technological problem to make a system of $N$ rings with the same radius. It is more difficult to make homogeneous system of rings with the same $T _{c}$. The resistive transition of a homogeneous system has a finite width because of thermal fluctuations \cite{Tink75}. A system is not quite homogeneous if its resistive transition is wider than the one of the ideal system. The width $\Delta T_{c}(0.1 \div 0.9R_{n}) \approx  0.02 \ K$ of the transition $R(T)$ of the rings system used in \cite {PLA2012} is approximately three times more than the ideal one $\approx  0.006 \ K$. The width $\Delta T_{c}(0.1 \div 0.9R_{n}) \approx  0.009 \ K$ of the system used in this work corresponds approximately the ideal one. Therefore the rectification efficiency can not be increased with the help of a more homogeneous system. 

One may increase the rectification efficiency of each ring $Eff _{Re,1} = V _{dc,i}/R _{n,1}I _{A}$. Here $R _{n,1} \approx 5400/667 \ \Omega \approx 8 \ \Omega $ is the resistance of one ring in the normal state. The dc voltage up to $V _{dc,i} \approx V _{dc,A}/667 \approx 1.5 \ nV$ is induced by the noise with $I _{A} \approx 50 \ nA$ on each ring and $Eff _{Re,1} \approx 0.004$. The $V _{dc,A}$ depends from the amplitude of the persistent current $I _{p,A}$ rather than from the noise amplitude $I _{A}$. The amplitude $I _{A}$ should exceed the critical current (4) in order the oscillations $V _{dc}(\Phi /\Phi _{0})$ could be observed. Therefore the value $I _{p,A}$ should be higher and the critical current $I _{c} = I _{c0}-2I _{p,A}2|n - \Phi /\Phi _{0}|$ should have a non-zero but small value in order the relation $V _{dc,i}/I _{A}$ could have highest value. The temperature dependencies  $I _{c0}(T)$ and  $I _{p,A}(T)$, Fig.2, correspond to the theory \cite{Tink75} according to which the depairing current  $I _{c0} = 2sn _{s}q\hbar /m\surd{3}\xi (T) $ and $I _{p,A} = 0.5\Phi _{0}/L _{k} = sn _{s}q0.5\hbar /mr$: $I _{c0}(T) \propto (1 - T/T _{c})^{3/2}$ and  $I _{p,A}(T) \propto (1 - T/T _{c})$ because the density of superconducting pairs $n _{s} = |\Psi |^{2} \propto (1 - T/T _{c})$ and the correlation length $\xi (T) = \xi (0) (1 - T/T _{c})^{-1/2}$. 

According to the relation $I _{c0}/ I _{p,A} = 4r/\surd{3}\xi (T)$ the amplitude of the persistent current equals the critical current at $1 - T/T _{c} \approx (\surd{3}\xi (0)/4r)^{2}$. Consequently one may expect that the rectification efficiency will increase with the radius $r$ decrease because $I _{p,A}(T) \propto (1 - T/T _{c})$. The noise with the amplitude $I _{A} \approx 50 \ nA $ can induce the dc voltage $V _{dc}(\Phi /\Phi _{0})$ in the temperature region down to $T/T _{c} \approx 0.98$ where $ I _{c}(\Phi = 0) \approx 250 \ nA$, Fig.2, because the persistent current reduces strongly the critical current in the temperature region $T/T _{c} > 0.98$. According to the theoretical dependence $ I _{p,A}(T)$, $I _{c0}(T)$ obtained on the base of our experimental result the equality $ I _{p,A}(T) \approx I _{c0}(T)$ is observed $1 - T/T _{c} \approx 0.01$. This value $(\surd{3}\xi (0)/4r)^{2} \approx 0.01$ corresponds to the ring radius $r \approx 0.5 \ \mu m$ and the correlation length of aluminium film $\xi (0) \approx 0.1 \ \mu m$ at $T = 0$ \cite{PLA2012}. One may expect that a noise with the amplitude $I _{A} < 50 \ nA $ will induce the dc voltage $V _{dc}(\Phi /\Phi _{0})$ in a system of aluminium rings with the diameter $2r \approx 0.3 \ \mu m$ in the temperature region $T/T _{c} \approx 0.9$ where the persistent current is much higher.

\section{Conclusion} 
The induction of the quantum oscillation of the dc voltage by a non-equilibrium noise was observed first on an asymmetric superconducting quantum interference device as far back as 1967 \cite{Physica1967}. The observations of the oscillations $V _{dc}(\Phi /\Phi _{0})$  with the amplitude up to $ V _{A} \approx 15 \ \mu V$ on the device with the resistance $R _{n} \approx 2 \ \Omega $ indicate that the amplitude of an uncontrollable noise in the measuring system of the authors \cite{Physica1967} exceeds strongly $ I _{A} \approx 20 \ \mu A$. This ratchet effect was rediscovered in 35 years when the oscillations $V _{dc}(\Phi /\Phi _{0})$ with the amplitude up to $ V _{A} \approx 1.2 \ \mu V$  were observed on a single asymmetric aluminium ring with $R _{n} \approx 15 \ \Omega $ \cite{PerMob2001}. The amplitude of a non-equilibrium noise should exceed  $3 \ \mu A$ in order to induce such oscillations. The visible oscillations $V _{dc}(\Phi /\Phi _{0})$ vanished after the reduction of the non-equilibrium noise with the help of screening and filtration. But the reduced noise could be detected with the help of bigger number of rings connected in series. The noise with $ I _{A} > 3 \ \mu A$ induced the visible oscillations with $V _{A} \approx 1.2 \ \mu V$ on a single asymmetric ring \cite{PerMob2001} whereas the visible oscillations with $V _{A} > 0.2 \ \mu V$ could not be observed on a system of 18 rings \cite{PCJETP07} when the non-equilibrium noise was reduced more than by a factor of ten, down to $ I _{A} \approx 0.25 \ \mu A$. This reduced noise could be detected with the help of a system of 110 rings connected in series \cite{Letter07}. The oscillations $V _{dc}(\Phi /\Phi _{0})$ with $V _{A} \leq  0.6 \ \mu V$ visible in \cite{Letter07} have become invisible after the additional filtration of the non-equilibrium noise with low-temperature $\pi $-filters and coaxial resistive twisted pairs. The noise reduced down to  $ I _{A} \approx 20 \ nA$ could be detected with the help of 1080 rings \cite {PLA2012}.  

Thus, the quantum oscillations $V _{dc}(\Phi /\Phi _{0})$ diminish rather than disappear when the non-equilibrium noise is reduced a thousand times, from  $ I _{A} > 20 \ \mu A$ in \cite{Physica1967} down to $ I _{A} \approx 20 \ nA$ in \cite {PLA2012}. The invisible oscillations $V _{dc}(\Phi /\Phi _{0})$ can by made visible due to the summing up of the dc voltage. Could the oscillations $V _{dc}(\Phi /\Phi _{0})$ disappear? Contemporary progress of nano-technology allows to answer on this question. It is not inconceivable that the oscillations $V _{dc}(\Phi /\Phi _{0})$ observed in \cite {PLA2012} on 1080 rings will be possible to make invisible with the help of an additional filtration of the non-equilibrium noise. But these oscillations will be possible to make visible again with the help of a system with a bigger number of rings. This possibility to increase the sensitivity of the noise detector has practical importance. It may also answer on a fundamental question: "Could the equilibrium noise induce the quantum oscillations of the dc voltage?" One can assume such possibility due to the observations of the persistent current in the fluctuation region near superconducting transition, where the ring resistance is not zero, Fig.2. The progress of nano-technology has allowed to observe the persistent current also in normal metal rings \cite{PCScien09,PCPRL09}. It is well known that the electric current $I$ circulating in a ring induces the potential difference $V = I(R _{h} - R _{l})/2$ on the ring halves having different resistance $R _{h} > R _{l}$. The progress of nano-technology allows to make a system of numerous asymmetric (with $R _{h} > R _{l}$) normal metal nano-rings connected in series. Measurements of such system can answer on the question: "Would the persistent current induce the potential difference?"

\end{document}